\documentclass{article}

\usepackage{amsmath}
\usepackage{xcolor}
\usepackage{bbold}
\usepackage{graphicx}
\usepackage{float}
\usepackage{algorithm}
\usepackage{algpseudocode}
\usepackage{mathtools}
\usepackage{ragged2e}
\usepackage{enumerate}
\usepackage{natbib}
\usepackage{subfig}
\usepackage{url}

\newtheorem{proposition}{Proposition}

\graphicspath{{./../archive/CSDA-251211/figs/}{./}} 

\title{\textcolor{black}{Gaussian copula correlation network analysis of mixed-type data using a semi-parametric pairwise likelihood with a multi-omics application}}

\newcommand{\affiliationflorence}{\addtocounter{footnote}{-1}\footnotemark}

\author{Ekaterina Tomilina%
  \thanks{Université Paris-Saclay, INRAE, MaIAGE, Jouy-en-Josas, France}
  \thanks{Université Paris-Saclay, INRAE, AgroParisTech, GABI, Jouy-en-Josas, France} 
  \and Florence Jaffrézic\affiliationflorence \and Gildas
  Mazo%
  \thanks{
    Université Paris-Saclay, INRAE, MaIAGE, Jouy-en-Josas,
    France (corresponding author). Email: \texttt{gildas.mazo@inrae.fr}}}

\begin{document}

\maketitle

\footnotetext{Attachments will appear as annexes in the electronic
  version of the manuscript.}

\begin{abstract}
  Reconstructing gene regulatory networks from large-scale
  heterogeneous data is a key challenge in biology. In multi-omics
  data analysis, networks based on pairwise statistical association
  measures remain popular, as they are easy to build and understand.
  In the presence of mixed-type (discrete and continuous) data,
  however, the choice of good association measures remains an
  important issue. It is proposed a novel approach based on the
  Gaussian copula, the parameters of which represent the links of the
  network.  Novel properties of the model are obtained to guide the
  interpretation of the network. To estimate the copula parameters,
  a semiparametric pairwise likelihood for mixed data was calculated. 
  An extensive simulation study showed that the proposed
  estimation procedure was able to accurately estimate the copula
  correlation matrix. The proposed methodology was also applied to a
  real ICGC dataset on breast cancer, and is implemented in a freely
  available R package $\texttt{heterocop}$.
\end{abstract}


\section{Introduction}
\label{s:intro}
The recent development of high-throughput sequencing technologies
provides access to a large amount of omics data of various types
(transcriptomics, proteomics, metabolomics, metagenomics,
epigenetics). To better understand the regulatory mechanisms that
underlie these data, the so-called correlation networks depict
statistical associations between pairs of measurements, both intra-
and inter-type.  A major statistical challenge underlying the
construction of correlation
networks is the heterogeneous nature of the data. Indeed,
RNA-seq data for instance are count data, whereas protein abundances
are continuous and mutation encoding is often binary. Existing methods
such as WGCNA~\citep{Langfelder2008} rely on Pearson's correlation
coefficient, and are therefore limited to linear relationships between
variables. Another possibility would be to base the construction of
the correlation network
 on Kendall's tau or Spearman's rho. However, these
coefficients are not well-adapted when at least one variable is
discrete~\citep{Neslehova2007,mesfioui2022,kendall_treatment_1945}. \textcolor{black}{For instance, Spearman's rho involving a binary random variable is bounded in absolute value by a quantity that depends on its parameter~\citep{mesfioui2022}.} \textcolor{black}{Other approaches such as Mutual Information computation~\citep{Margolin2006, Qiu2009, Ross2014} are available for mixed variables but cannot distinguish between positive and negative associations and do not rely on a statistical model. Indeed,}
the lack of an underlying model may be an impediment to further
statistical investigation, for instance the simulation of new data or
the inclusion of covariables in the experimental design. The goal of
this paper is therefore to propose a novel approach to correlation
network analysis of mixed-type data, based on the Gaussian copula.

The Gaussian copula model relies on the assumption that the observed variables are transformations of a hidden Gaussian vector, and enables to link their joint cumulative distribution function (CDF) to a Gaussian CDF while preserving their marginal distributions. The Gaussian copula model corresponds to the Nonparanormal distribution in the continuous case~\citep{Liu2009}, but can be extended to discrete and mixed variables. With this approach, for instance, it is possible to build a joint CDF for a Poisson, a Negative Binomial and a Gamma random variable, which enables to deal with biological data of various nature.

As biological data do not perfectly follow a pre-defined distribution, semi-parametric approaches have been proposed in~\citep{Fan2017,Yoon2019,yoon_sparse_2020,Dey2022,Gobler2024}. \textcolor{black}{These methods rely on a pairwise correspondance---called a bridge function---between the copula correlation coefficient and some statistic of the observed variables. However those bridge functions need to be constructed on a case-by-case basis~\citep{Dey2022,Gobler2024} and can be computationally expensive or unstable~\citep{yoon_fast_2021}.}

We introduce a more direct, likelihood-based approach. We provide an explicit expression of the pseudo-likelihood in the mixed case of continuous and discrete variables, and give a detailed theoretical proof of its calculation. As multi-omics data are often high-dimensional, a pairwise likelihood estimator is built. In order to avoid assumptions on the distribution of the marginals, we estimate the CDFs empirically. 

We show the equivalence between the presence of a block-wise diagonal structure in the copula correlation matrix and block-wise mutual independence in the observed data. We characterize the lower and upper extreme values of the copula parameter in terms of the observed data when a Bernoulli distribution is involved.
This provides an interpretation of the copula correlation coefficients in terms of association relationships between the observed variables. The performance of the proposed method is illustrated in an extensive simulation study. An application to a real ICGC~\citep{ICGC} dataset containing tumoral samples of women affected by breast cancer is carried out.

The rest of the paper is as follows. Section~\ref{s:model} presents the model and its dependence properties.  The estimation method is given in Section~\ref{s:inference}. Section~\ref{s:simu} presents the simulation studies and Section~\ref{s:app} the real data analysis. A discussion section closes the paper.

\section{The model}
\label{s:model}

Let $(X_1, \dots, X_d)$ be a random vector with cumulative distribution function (CDF) given by
\begin{equation}
\begin{split}
    F(x_1, ..., x_d)&=C_{\Sigma}(F_1(x_1), \dots, F_d(x_d))\\
    &\equiv\Phi_\Sigma(\Phi^{-1}(F_1(x_1)), ..., \Phi^{-1}(F_d(x_d))),
    \end{split}
     \label{eqn:model}
\end{equation}
where $F_1, \dots, F_d$ denote the marginal CDFs of the variables $X_1, \dots, X_d$, $C_{\Sigma}$ denotes the Gaussian copula parameterized by the correlation matrix $\Sigma$, $\Phi_\Sigma$ the centered Gaussian multivariate CDF of correlation matrix $\Sigma$, and $\Phi^{-1}$ the inverse of the standard Normal CDF $\Phi$. It can be checked that the right-hand side of~(\ref{eqn:model}) indeed is a well-defined CDF with marginals $F_1,\dots,F_d$~\citep{Sklar1973, Nelsen2007}.
One can note that model~(\ref{eqn:model}) corresponds to a latent Gaussian variable structure. \textcolor{black}{Indeed, each $X_j$ can be expressed as $X_j=F_j^\leftarrow(\Phi(Z_j))$, where $(Z_1, ...., Z_d)\sim \mathcal{N}(0, \Sigma)$ is a centered Gaussian vector such that each $Z_j$ is standardized with variance $1$ (so $\Sigma$ is a correlation matrix).} Note that $F^{\leftarrow}_j$ denotes the generalized inverse function of $F_j$, that is, $F^{\leftarrow}_j(u)=\inf\{x: F_j(x) \geq u \}$. \textcolor{black}{We have chosen the notation $F_j^{\leftarrow}$ to emphasize that  for discrete variables the distribution function is not invertible. Note that for continuous variables, $F_j^{\leftarrow}$ coincides with the usual inverse $F_j^{-1}$.}
With model~(\ref{eqn:model}) we do not assume  that the observed variables $X_1,\dots,X_d$ are Gaussian. Only the latent variables $Z_1,\dots,Z_d$ are. In model~(\ref{eqn:model}) the marginal distributions $F_1,\dots,F_d$ of the observed variables $X_1,\dots,X_d$ are arbitrary. In particular, there can be a mix of continuous and discrete variables.
Model~(\ref{eqn:model})  also provides us with an explicit expression of the joint CDF of the variables as a function of their marginal CDFs and thus enables us to see how the distribution of each variable impacts the joint distribution. Note that when all the variables are continuous, model~(\ref{eqn:model}) corresponds to the Nonparanormal distribution defined in~\citet{Liu2009}.

\subsection{Joint density}
\label{sec:jointdensity}

An expression of the multivariate density can be derived from model~(\ref{eqn:model}). Below, we say that a random variable is continuous if its CDF is increasing, and discrete if its CDF has a countable support. 

\begin{proposition}
\label{prop:density}
Without loss of generality, suppose that the first $p$ variables are continuous and that the remaining $d-p$ are discrete. Then the  multivariate density of~(\ref{eqn:model}) can be written as
\begin{multline}
  f(x_1, \dots, x_d)=\left(\prod_{j=1}^{p}f_j(x_j)\right)\times\\
\sum_{j_{p+1}=0}^{1}\cdots\sum_{j_d=0}^{1}(-1)^{j_{p+1}+\cdots+j_d}
   C_{\Sigma}^{p}(F_1(x_1), \dots, F_p(x_{p}),u_{p+1,j_{p+1}},\dots,u_{d,j_d}),
  \label{eqn:density}
  \end{multline}
  where $f_j$ denotes the density of $X_j$, $u_{j,0}=F_j(x_j)$ and $u_{j,1}=F_j(x_j-)$, $x_j-$ denotes the previous point from $x_j$ in the ordered support of $F_j$, and 
 $C_\Sigma^p$ denotes the derivative of the copula with respect to the $p$ continuous marginal CDFs, that is $C_\Sigma^p(u_1,\dots,u_d)=\partial^pC_\Sigma(u_1, \dots, u_d)/\partial u_1 \cdots \partial u_p$. If $x_j$ is the least point (if there is one), we set by convention that $F_j(x_j-)=0$.
Also by convention we set that if $p=d$ then the second factor in the right-hand side of~(\ref{eqn:density}) is replaced by $c_\Sigma(F_1(x_1),\dots,F_p(x_p))$, where $c_\Sigma(u_1,\dots,u_p) = C^p_\Sigma(u_1,\dots,u_p)$ is the density of $C_\Sigma$. If $p=0$, the first factor in~(\ref{eqn:density})  is replaced by 1 and $C_\Sigma^p(u_1,\dots,u_d) = C_\Sigma(u_1,\dots,u_d)$.
\end{proposition}

The formula~(\ref{eqn:density}) appears in~\citet{Song2007} without proof. A proof of Proposition~\ref{prop:density} is given in Section~C.1 of the Supplementary material. 

\subsection{Dependence properties}
\label{sec:depproperties}

Having an expression of the multivariate density in equation~(\ref{eqn:density}) enables us to study the (in)dependence relationships between $X_1, \dots, X_d$.

\subsubsection{Multivariate dependence properties}

\begin{proposition}
  \label{prop:block_wise}
  Let $G_1, \dots, G_k$ be a partition of $D=\{1, \dots d\}$, and denote $X_G=(X_j : j \in G)$ for $G \subset D$. Then, $X_{G_1}, \dots, X_{G_k}$ are mutually independent if and only if $\Sigma$ is a block matrix of the form
  \begin{equation*}
    \Sigma=\begin{pmatrix}
      \Sigma_1 & 0 & ... & 0\\
      0 & \Sigma_2 & ... & 0\\
      0 & 0 & ... & 0 \\
      0 & 0 & 0 & \Sigma_k
    \end{pmatrix},
  \end{equation*}
  where each $\Sigma_i$ is a block of size $|G_i| \times |G_i|$.
\end{proposition}

A proof of Proposition~\ref{prop:block_wise} can be found in Section~C.2 of the Supplementary material. We see that the correlation matrix of the copula encodes mutual independencies between groups of variables.

\subsubsection{Bivariate dependence properties}

Let $X_1$ and $X_2$ be a pair of variables distributed according to the Gaussian copula model~(\ref{eqn:model}) with
\begin{equation*}
  \Sigma =
  \begin{pmatrix}
    1&\rho\\\rho&1
  \end{pmatrix}.
\end{equation*}
In this case the copula is simply denoted by $C_\rho$. Using~(\ref{eqn:model}), it is easy to see that the density $c_\rho$ of $C_\rho$, that is, $c_\rho(u,v) = \partial^2 C_\rho(u,v)/\partial u\partial v$, $0<u,v<1$, is given by
\begin{equation*}
c_\rho(u,v)=\dfrac{1}{\sqrt{1-\rho^2}} \exp\left(\dfrac{2\rho\Phi^{-1}(u)\Phi^{-1}(v)-\rho^2(\Phi^{-1}(u)^2+\Phi^{-1}(v)^2)}{2(1-\rho^2)}\right).
\end{equation*}

By definition, the parameter $\rho$  measures the correlation between the latent Gaussian variables, but how can it be interpreted for the observed variables? By taking $d=2$ in  Proposition~\ref{prop:block_wise} we see that $\rho=0$ if and only if $X_1$ and $X_2$ are independent. We shall see that the lower and upper extreme values of $\rho$ can also be characterized in terms of the observed variables when the discrete variables follow a Bernoulli distribution. Remember that $X_1$ and $X_2$ are said to be \emph{comonotonic} if one of them is almost surely an increasing function of the other, and \emph{countermonotonic} if they are almost surely a decreasing function of each other~\citep{Nelsen2007}.

\begin{proposition}\label{prop:characterization}
  Suppose that one of the three cases below holds:
    \begin{enumerate}[(i)]
    \item\label{item:increas-increas} $X_1$ and $X_2$ are continuous;
    \item\label{item:Ber-increas} $X_1\sim\mathcal{B}(p_1)$, $0<p_1<1$, and $X_2$ continuous;
    \item\label{item:Ber-Ber} $X_1\sim \mathcal{B}(p_1)$, 
      $X_2\sim \mathcal{B}(p_2)$, $0 < p_1 \le p_2 < 1$ and $p_1 + p_2 \ge 1$.
    \end{enumerate}
Then  
  \begin{equation*}
    \rho = 1 \text{ iff }\left\{
      \begin{array}{rl}
        (X_1,X_2) \text{ is comonotonic }&\text{ case~(i);}\\
        (X_1,\mathbf{1}_{\{X_2>F_2^{-1}(1-p_1)\}}) \text{ is comonotonic }&\text{ case~(ii) ;}\\
        X_1\le X_2 & \text{ case~(iii)};
      \end{array}\right.
  \end{equation*}
  and
  \begin{equation*}
    \rho = -1 \text{ iff }\left\{
      \begin{array}{rl}
        (X_1,X_2) \text{ is countermonotonic }&\text{ case~(i);}\\
        (X_1,\mathbf{1}_{\{X_2>F_2^{-1}(p_1)\}}) \text{ is countermonotonic }&\text{ case~(ii) ;}\\
        X_1 +X_2 > 0 & \text{ case~(iii)}.
      \end{array}\right.
  \end{equation*}
\end{proposition}

A proof of Proposition~\ref{prop:characterization} is given  in Section~C.3  of the Supplementary material. In case~(ii) for $\rho=1$, the variable $X_2$ exceeds a certain threshold only if $X_1=1$. A similar pattern holds for $\rho=-1$. In case~(iii), $\rho=1$ indicates that $X_1$ is dominated by $X_2$, and $\rho=-1$ indicates that at least one of the variables has to be non-null. For example, if $X_1$ and $X_2$ encode the presence of two mutations, then $\rho=1$ indicates that presence of the first mutation implies presence of the second. A visual representation of Proposition~\ref{prop:characterization} is depicted in Figures~S1 ($\rho=1$) and~S2 ($\rho=-1$) of the Supplementary material.

\section{Estimation of $\Sigma$}
\label{s:inference}
Let $X^i=(X_1^i,\dots,X_d^i)$, $i=1,\dots,n$, be
$n$ i.i.d. observations in  $\mathbb{R}^d$ drawn from the distribution defined in model~(\ref{eqn:model}).  As it is often the case, we suppose that for all $j$ in $\{1, \dots, d\}$, we have no information regarding the marginal distributions $F_j$ which are replaced by the empirical distributions $$\hat{F}_j(x)=\dfrac{1}{n}\sum_{i=1}^n \mathbb{1}(X_j^i \leq x),$$ where $\mathbb{1}$ denotes the indicator function. Hence, our estimation is performed in a semi-parametric framework. In a high dimensional setting, computing the full multivariate density has a high computational cost. We therefore propose to estimate the copula correlation matrix $\Sigma$ by extending the pairwise  maximum likelihood estimator~\citep{Mazo2022} to mixed and non-parametric marginals. In other words, we compute
\begin{equation}
  \hat{\Sigma}=\underset{\Sigma}{\arg\max}\dfrac{1}{n}\sum_{i=1}^n \sum_{j<j'} \log \hat{f}_{jj'}(X_j^i, X_{j'}^i, \rho_{jj'}).
   \label{eqn:estimator}
 \end{equation}
 In the expression above, $\rho_{jj'}$ denotes the element of $\Sigma$ at the $j$th row and $j'$th column and $\hat f_{jj'}$ denotes an estimate of the density of the bivariate marginal CDF corresponding to $(X_j,X_{j'})$ with respect to $\lambda\otimes\lambda$ if both variables are continuous, $\mu\otimes\mu$ if both variables are discrete, and $\lambda\otimes\mu$ measure if $X_j$ is continuous and $X_{j'}$ is discrete, with $\lambda$ the Lebesgue measure and $\mu$ the counting measure.
 Above we said that $\hat f_{jj'}$ is an \emph{estimate} of $f_{jj'}$, the density of $(X_j,X_{j'})$. Indeed, as we shall see below, the density $f_{jj'}$ depends on the marginal CDFs $F_j$ and $F_{j'}$. But since we substitute the empirical CDFs $\hat F_j$ and $\hat F_{j'}$ for $F_j$ and $F_{j'}$, the resulting function $\hat{f}_{jj'}(\cdot,\cdot;\rho_{jj'})$ is only an estimate of the true density $f_{jj'}(\cdot,\cdot;\rho_{jj'})$.

 The formulas of the densities $f_{jj'}$ in the three cases ($X_j$ and $X_{j'}$ continuous, $X_j$ continuous and $X_{j'}$ discrete, $X_{j}$ and $X_{j'}$ discrete) are given next. 
 Rewrite
 \begin{equation*}
   C_{\rho_{jj'}}(u,v) = C_{\Sigma}(1,\dots,1,u,1,\dots,1,v,1,\dots,1),
 \end{equation*}
 $u$ and $v$ at the $j$th and $j'$th positions, respectively,
 so that  the bivariate CDF of $(X_j,X_{j'})$ is given by $C_{\rho_{jj'}}(F_j(x_j),F_{j'}(x_{j'}))$.
Let $c_{\rho_{jj'}}(u,v)$ denote the density of $C_{\rho_{jj'}}(u,v)$, that is, $c_{\rho_{jj'}}(u,v) = \partial^2 C_{\rho_{jj'}}(u,v)/\partial u\partial v$, $0<u,v<1$, $-1<\rho_{jj'}<1$.
Let $f_j$ be the marginal density of variable $X_j$. If $X_j$ and $X_{j'}$ are continuous, then $f_{jj'}$ can be expressed as
\begin{equation*}
    f_{jj'}(x_j, x_{j'})=f_j(x_j)f_{j'}(x_{j'})\times c_{\rho_{jj'}}(F_j(x_j),F_{j'}(x_{j'})).
\end{equation*}
If both variables are discrete, then the density takes the following form:
\begin{equation*}
\begin{split}
  f_{jj'}(x_j,x_{j'})=\mathbb{P}(X_j=x_j, X_{j'}=x_{j'})&=C_{\rho_{jj'}}(F_j(x_j),F_{j'}(x_{j'}))\\
  &\text{ }\quad+ C_{\rho_{jj'}}(F_j(x_j-),F_{j'}(x_{j'}-))\\
  &\text{ }\quad- C_{\rho_{jj'}}(F_j(x_j-),F_{j'}(x_{j'}))\\
  &\text{ }\quad-C_{\rho_{jj'}}(F_j(x_j),F_{j'}(x_{j'}-)).
  \end{split}
\end{equation*}
Finally, if $X_j$ is continuous and $X_{j'}$ is discrete, then we get the following form:
  \begin{equation*}
   f_{jj'}(x_j, x_{j'})=f_j(x_j)\int_{F_{j'}(x_{j'}-)}^{F_{j'}(x_{j'})}c_{\rho_{jj'}}(F_j(x_j), v)\mathrm{d}v.
\end{equation*}
The estimated density $\hat f_{jj'}$ is obtained by substituting $\hat F_j$ and $\hat F_{j'}$ for $F_j$ and $F_{j'}$, respectively, in the formulas above.

\section{Simulations}
\label{s:simu}
The goal of this simulation study was to illustrate several properties of the proposed copula model and estimation procedure. We first considered the bivariate  case. Then, we extended our estimation to a high-dimensional setting. \textcolor{black}{Finally, execution time was investigated.} The simulations from this section were run with our $\texttt{heterocop}$ R package available on CRAN. 

\subsection{Simulation study in the bivariate case}
We simulated four variables with a joint cumulative distribution function corresponding to a  Gaussian copula as in model~(\ref{eqn:model}) and with marginals detailed below:
\begin{itemize}
    \item a Poisson distribution $\mathcal{P}(1)$ of mean and variance 1;
    \item a Negative Binomial distribution, denoted $NB(1,0.5)$, where $1$ is the number of successful trials and $0.5$ is the probability of success;
    \item \textcolor{black}{a standard normal distribution $\mathcal{N}(0,1)$};
    \item a Bernoulli distribution $\mathcal{B}(0.5)$ of mean $0.5$.
\end{itemize}

\begin{figure}[H]
  \includegraphics[width=\linewidth]{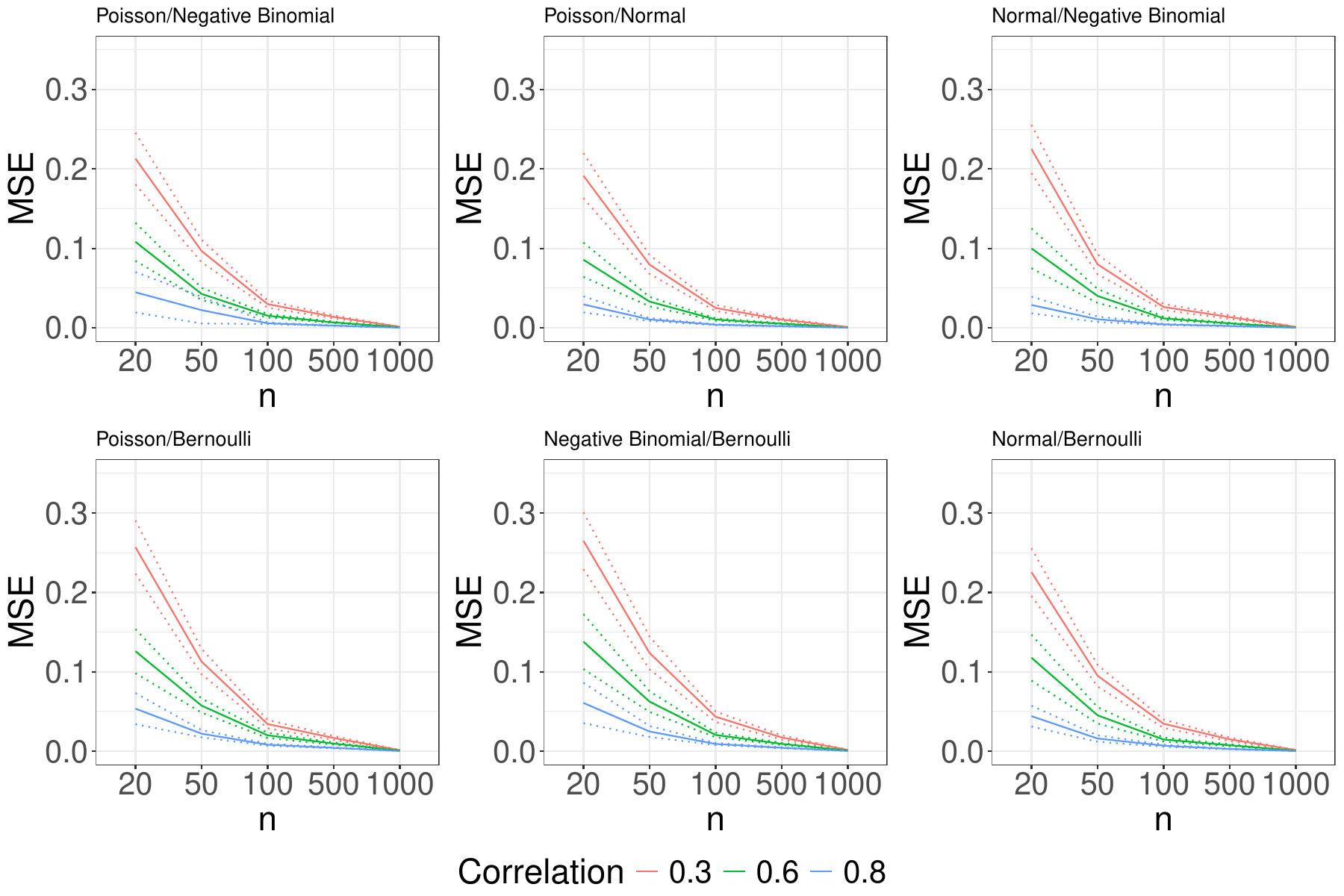}
  \caption{Estimated MSE of the estimator $\hat{\rho}_n$ from $N=500$
    simulations for each sample size $n$, value of the copula
    coefficient $\rho$, and type of marginal distributions (plain
    lines). Each pair of dotted lines denotes 95\% confidence
    intervals constructed from the 500 simulations for the true
    MSE. This offers an evaluation of the uncertainty about each MSE's
    estimate caused by the randomness of the simulations. See
    Section~G of the Supplementary
    material for details about the calculation of these confidence
    intervals.}
  \label{fig:mse}
\end{figure}

The four variables make 6 pairs, studied separately.
Let $\rho$ denote the copula parameter of the pair considered and $\hat{\rho}$ its estimate obtained from~(\ref{eqn:estimator}).
The Mean Squared Error (MSE) of $\hat{\rho}$ is defined as:
       $\text{MSE}(\hat{\rho})=\mathbb{E}[(\hat{\rho}-\rho)^2]$.
The MSE can be decomposed into the sum of the variance and the squared bias of $\hat{\rho}$ as follows:
\begin{equation*}
    \begin{split}
       \text{MSE}(\hat{\rho})&=\mathbb{E}[(\hat{\rho}-\rho)^2] = \underbrace{\mathbb{E}[\hat{\rho}^2]-\mathbb{E}[\hat{\rho}]^2}_{Var(\hat{\rho})} + \underbrace{(\mathbb{E}[\hat{\rho}-\rho])^2}_{Bias(\hat{\rho})^2}.
    \end{split}
\end{equation*}
For each of the 6 pairs, the MSE, variance and squared bias of our estimator were empirically estimated by running $N=500$ simulations for different sample sizes $n=20, 50, 100, 500, 1000$ and copula coefficients $\rho=0.3, 0.6, 0.8$. The results are depicted in Figure~\ref{fig:mse}. 

Figures~S3 and S4 of the Supplementary material represent the evolution of the variance and of the squared bias depending on the sample size. One can see that the variance of the estimators decreases to zero as the sample size increases. It is also interesting to note that it is higher for lower values of the correlation coefficient ($\rho=0.3$) than for the higher ones ($\rho=0.8)$. The variances do not seem to be impacted by the types of the variables, and a similar pattern is observed for all the distributions considered here. It can be noticed that the squared biases are all very close to zero. Although slightly higher in the discrete/discrete case for $n=20$, they remain extremely low and do not significantly differ from zero as soon as the sample size exceeds 50.

\textcolor{black}{The same simulations have been run for negative values $\rho=0, -0.3, -0.6, -0.8$. Section~E of the Supplementary material shows that the results are similar to the case of positive values: the bias and standard deviation of $\hat\rho$ also tend to zero as $n$ increases.}

We compared our semi-parametric approach with a fully parametric one in which  the parametric families of the marginal distributions are  known. We considered the case of the Normal and Negative Binomial distributions, with parameters specified above.
The copula correlation coefficient was now estimated in a fully parametric way, i.e. the parameters of the  marginals were estimated  by maximum likelihood in a first step (each marginal separately) and the copula parameter was estimated in a second step from the likelihood with the estimated parametric marginals plugged in. Figure~S5 in the Supplementary material presents the variances and squared biases of both the parametric and semi-parametric estimates, for $N=500$ simulations. The variances were found to be slightly higher for our semi-parametric estimator for low sample sizes of 20 and 50, but quite similar otherwise. Both the parametric and semiparametric estimators have a negligible bias, compared to the variance.

In real data analyses, the parametric families of the marginal distributions is rarely known. We therefore
 assessed the robustness of our semi-parametric method against miss-specification of the marginal distributions.
We simulated the data as previously but estimated the marginal parameters assuming  a Poisson distribution instead of a Negative Binomial one, a common situation in genomics.
Figure~S6 of the Supplementary material shows  that the estimates of $\rho$ obtained by the fully parametric approach are biased while our semi-parametric method remains robust. Our proposed approach will therefore be useful for practical applications when the parametric distribution of the data cannot be specified.

\begin{figure}[H]
  \includegraphics[width=\linewidth]{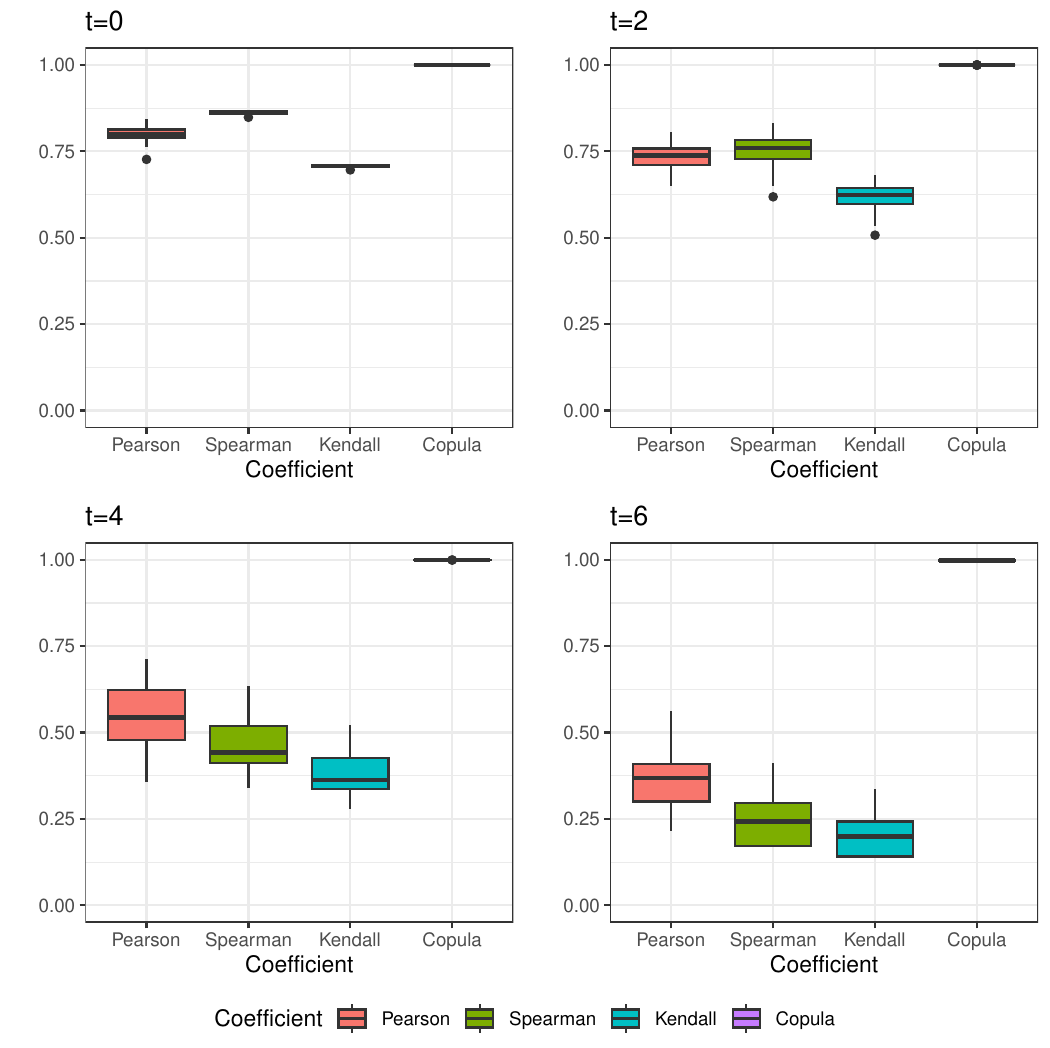}
  \caption{Boxplot of 50 estimates, each based on a sample of size
    100, of each type of coefficient (Pearson, Spearman, Kendall,
    Copula) and for different thresholds $t$.}
  \label{fig:corr_coefs}
\end{figure}

Finally, we assessed the ability of the copula correlation coefficient
to capture complex dependence relationships.  Let
$X_1 \sim \mathcal{N}(0,3)$ and $X_2 = \mathbb{1}_{\{X_1 \ge t\}}$,
where $t \in \mathbb R$ is some fixed threshold. It is shown in
Section~D of the supplementary material that the
random vector $(X_1,X_2)$ belongs to model~(\ref{eqn:model}) with
copula correlation $\rho=1$. As shown in Figure~\ref{fig:corr_coefs},
the higher the threshold, the less the Pearson, Spearman or
Kendall coefficients are able to capture the dependence relationship. (Exact
numerical values of Pearson's $\rho^P$ and Spearman's $\rho^S$ are
0.79, 0.62, 0.27, 0.06, and 0.87, 0.57, 0.18, 0.03, respectively; see
Section~D of the supplementary material for the
calculations.) The proposed copula correlation estimation seems
therefore more robust when binary variables have to be analyzed,
especially in the case of rare events as observed in mutation data for
example, as presented in the next section.

\subsection{Simulation study in the high-dimensional case}

\subsubsection{Simulation protocol}
Five different sample sizes were considered $n=20, 50, 100, 500, 1000$, for $d=30$ and $d=300$ variables. In each case, one third of the variables were distributed according to a $\mathcal{N}(0,1)$, one third were $NB(1,\frac{1}{2})$ and the last ones were $\mathcal{B}(\frac{1}{2})$. 
Two structures were considered for the copula correlation matrix.
The first is a block-diagonal structure as specified below:
\[
  \Sigma_{blocks}= \left(\begin{array}{@{}c@{\quad}c@{\quad}c@{\quad}c@{\quad}c}
    \underbracket[.4pt]{\text{{0.8}}}_{7\times 7} & \text{0} &\text{0} & \text{0} & \text{0} 
    \\
    \text{0} & \underbracket[.4pt]{\text{{0.6}}}_{10\times 10} & \text{ 0} & \text{0} & \text{0} 
    \\
    \text{0}& \text{0}  & \underbracket[.4pt]{\text{{0.5}}}_{2\times 2} & \text{0} & \text{0} \\
    \text{0} & \text{0}& \text{0} &\underbracket[.4pt]{\text{{0.7}}}_{6\times 6} & \text{0} \\
    \text{0} & \text{0} & \text{0} & \text{0}&\underbracket[.4pt]{\text{{0.3}}}_{5\times 5} 
  \end{array}\right).
\]
In this matrix, the order of the variables was randomly defined to have blocks of correlated variables of different types. For $d=300$ variables, the size of each block was multiplied by 10. 
The second structure is a sparse structure. A matrix $\Sigma$ is generated through a modified Cholesky decomposition as described in Algorithm~\ref{alg:cap}. 

\begin{algorithm}[H]
  \caption{Simulation of a positive definite sparse matrix $\Sigma$}
  \label{alg:cap}
\begin{algorithmic}
\Require $\gamma \in [0,1]$, $m > 0$.
\State Define a $m\times m$ matrix of zeroes $\Sigma$.
\State Simulate $\frac{m(m-1)}{2}$ uniform $\mathcal{U}(0.3,1)$ coefficients.
\State Randomly set a proportion $\gamma$ of coefficients to 0.
\State Fill the upper triangular part of $\Sigma$ with the coefficients.
\State Do $\Sigma \leftarrow \Sigma^T\Sigma$.
\State Do $\Sigma_{ij} \leftarrow \dfrac{\Sigma_{ij}}{\sqrt{\Sigma_{ii}}\sqrt{\Sigma_{jj}}}$.  
\State Return $\Sigma$.
\end{algorithmic}
\end{algorithm}

By varying $\gamma$ in Algorithm~\ref{alg:cap}, we can generate matrices with different proportions of zeroes. We let $\gamma_F$ denote the obtained proportion of zeroes of the final matrix $\Sigma$. We call  $\gamma_F$ the  sparsity coefficient.
Regarding the sparsity of the correlation matrix, we have considered a final proportion $\gamma_F$ of null coefficients of around $20\%$, $50\%$ and $80\%$, by empirically setting the $\gamma$ parameter at 0.61, 0.79 and 0.91. The simulated matrices were denoted $\Sigma_{0.2}, \Sigma_{0.5}$, and $\Sigma_{0.8}$.
For each correlation matrix, simulations were run $N=500$ times.

\subsubsection{Numerical results}

Results are first presented for $d=30$ variables. 
The estimation accuracy was evaluated using the normalized Root Mean Squared Error (RMSE) and the normalized Mean Absolute Error (MAE) calculated as follows for a $d\times d$ matrix $\Sigma$:
\begin{equation}
    \begin{split}
       \mathrm{RMSE}(\hat{\Sigma})&=\frac{1}{N}\sum_{k=1}^N\sqrt{\frac{1}{d(d-1)}\sum_{1 \leq i\neq j\leq d}(\hat{\Sigma}_{ij}^k-\Sigma_{ij})^2}, \nonumber
    \end{split}
\end{equation}
\begin{equation}
    \begin{split}
       \mathrm{MAE}(\hat{\Sigma})&=\frac{1}{N}\sum_{k=1}^N\frac{1}{d(d-1)}\sum_{1 \leq i\neq j\leq d}|\hat{\Sigma}_{ij}^k-\Sigma_{ij}|, \nonumber
    \end{split}
\end{equation}
where $\hat{\Sigma}^k$ corresponds to the $k$th estimation of $\Sigma$. Note that the same $\Sigma$ was kept for the $N=500$ simulations.
As shown in Table~S1a from the Supplementary material, the normalized RMSE decreases as the sample size increases. It does not exceed $20\%$ for samples larger than $n=50$ and remains below $5\%$ for sample sizes greater than $n=500$. It seems to be  robust to the specification of the correlation structure and the amount of sparsity in the matrix.
Similarly to the normalized RMSE, the normalized MAE values given in Table~S1b from the Supplementary material decrease when the sample size increases, and remain below $5\%$ for sample sizes larger than $n=500$. The normalized MAE also seems robust to the structure of the correlation matrix and its sparsity. 

The same metrics were also evaluated in a higher-dimensional setting, for $d=300$ variables. In order to reduce computational time, we chose to study only a matrix of sparsity close to 0.8 and four different sample sizes $n=20, 50, 100, 500$. Table~S2 from the Supplementary material shows the obtained normalized RMSE and MAE also averaged over $N=500$ repetitions. The proposed estimation procedure was found to be robust to an increase of the number of variables. Normalized RMSE and MAE values were indeed  close to the values previously obtained with 30 variables, even for a small sample size. This result is promising for applying the proposed method to the analysis of real-life examples.

In the perspective of applying the proposed procedure to construct biological networks, we evaluate its ability to 
discriminate between small and large values of the copula correlation coefficient. 
Given a fixed threshold $t \in [0,1]$, a copula correlation coefficient estimate $\hat\rho$ is classified as belonging to the first group if $\textcolor{black}{|}\hat\rho\textcolor{black}{|}<t$, and as belonging to the second otherwise. By an abuse of language, we call the estimates classified into the first group the predicted zeroes, and those classified into the second group the predicted non-zeroes. Threshold $t$ was here arbitrarily set to 0.3.

The sensitivity to the identification of the non-zeroes, also known as true positive rate, and its specificity in the detection, also known as the true negative rate, were measured.
Let TP and FN denote the detected non-zeroes and detected zeroes, respectively, among the real non-zeroes. Similarly, let TN and FP denote the detected zeroes and detected non-zeroes among the real zeroes. 
The true positive rate (TPR) is equal to the proportion of detected non-zeroes among the real non-zeroes, that is, TPR=TP/(TP+FN). The true negative rate (TNR) is equal to the proportion of detected zeroes among the true zeroes, that is, TNR=TN/(TN+FP). The false negative rate (FNR) is defined as the proportion of detected non-zeroes among the real zeroes, that is, FNR=1-TNR. The false positive rate (FPR) is the proportion of detected zeroes among the real non-zeroes, that is, FPR = 1-TPR. A contingency table is available in Table~S3 of the Supplementary material for visual aid.

The Receiver Operating Characteristic (ROC) is a measure of global performance of a given classification rule, or classifier. It is a plot of  the TPR against the FPR for each value of $t$. For instance, when $t=0$, all the estimated coefficients are classified as non-zeroes and hence TPR=1, FNR=1.  When $t=1$, all the estimated coefficients are classified as zeroes and TPR=0, FNR=0. The AUC, Area Under Curve criterion enables us to quantify the performance of the classifier by evaluating the area under the ROC curve. The closer it is to 1, the better the performance.

\begin{figure}[H]
\centering
\includegraphics[width=\linewidth]{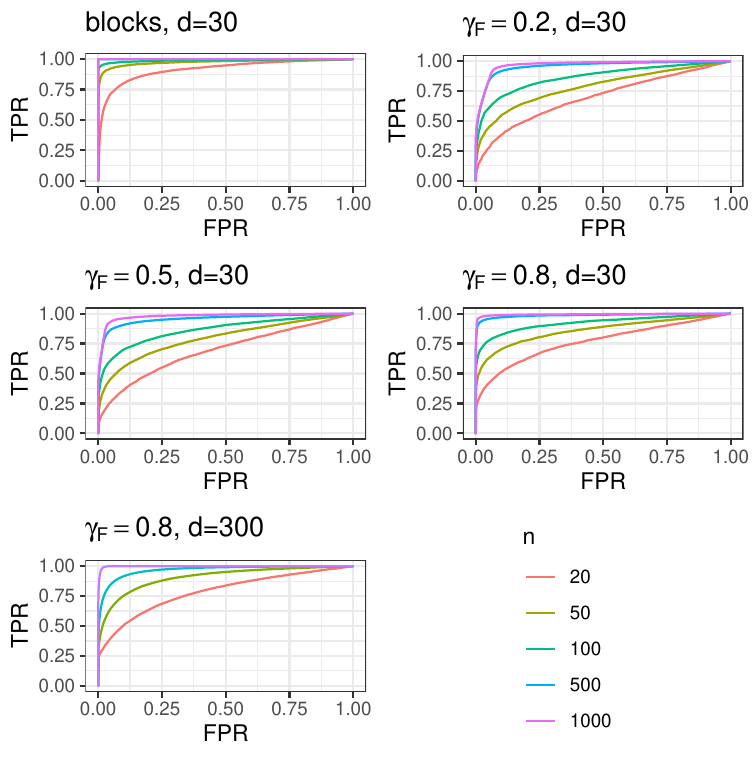}
\caption{Average ROC curves for $N=500$ simulations for the classification of the estimates of the copula correlation coefficients for different sample sizes. Four different sparse matrices were considered for $d=30$ variables (block-wise and sparse matrices with sparsity $\gamma_F=0.2, 0.5, 0.8$). A matrix with sparsity $\gamma_F=0.8$ was considered for $d=300$ variables.}
\label{fig:ROC}
\end{figure}

The ROC curves are presented in Figure~\ref{fig:ROC} for the four correlation structures considered for $d=30$ variables and each sample size, after averaging over $N=500$ simulations.
Figure~\ref{fig:ROC} also shows the results for $d=300$ variables for a matrix of 0.8 sparsity and Table~S4 from the Supplementary material sums up the AUC values in each case. As expected, the AUC values increase with the sample size. They are already  good for a low sample size of 20, close to 0.8 even for $d=300$ variables,  increase to around 0.9 for a sample size of 50, and close to 1 even for a sample size of 100. It can also be noticed that the accuracy is improved for a sparser correlation structure, which is often the case of interest in the context of biological network inference.

\subsection{\textcolor{black}{Scalability}}
\label{s:scalability}

\begin{figure}[H]
\centering
\includegraphics[width=\linewidth]{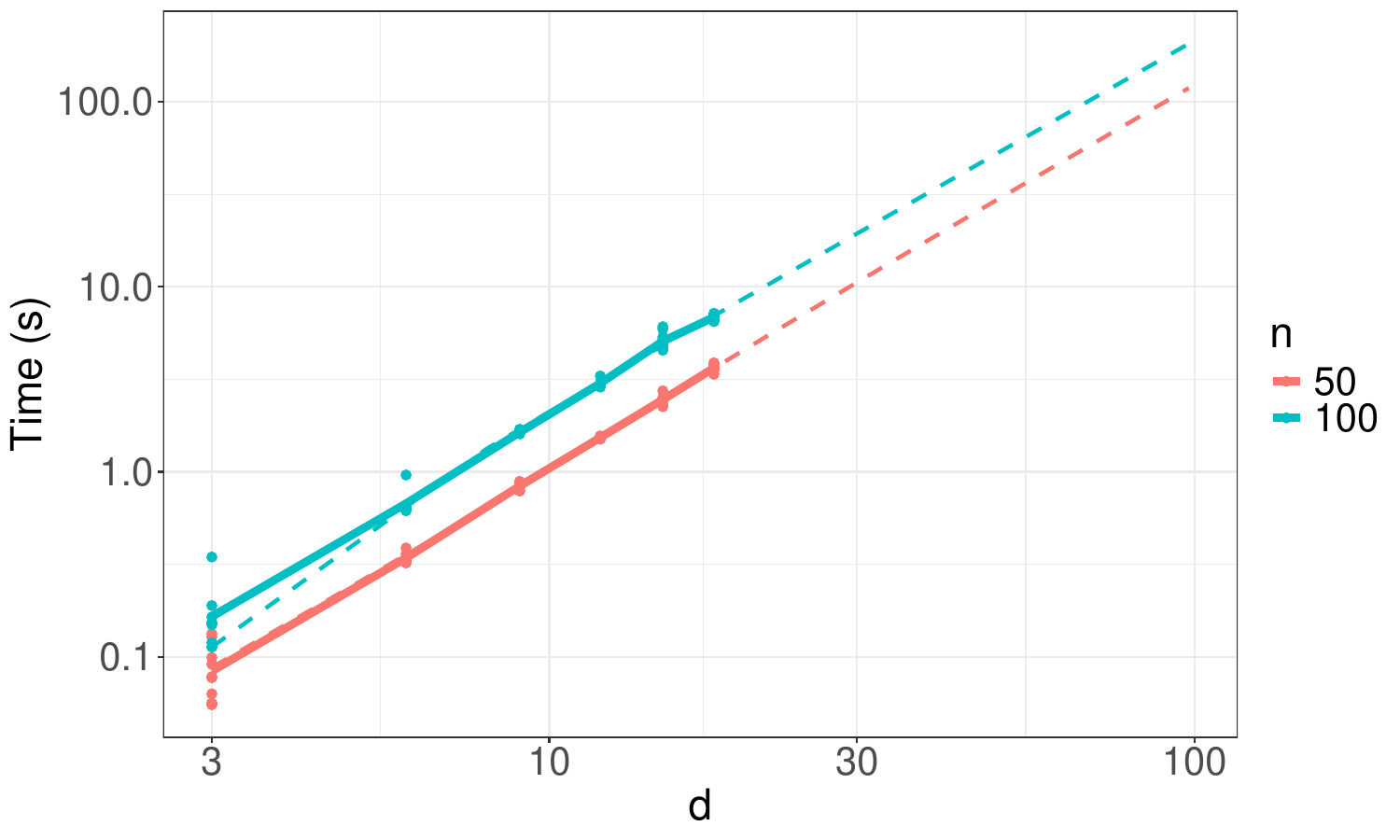}
\caption{Computational time ($\log_{10}-$scale on both axes) for the PPMLE for various values of $n$, predicted for higher values of $d$ (dotted lines) via a fitted second degree $\texttt{lm}$ on $d\in\{3,6,9,12,15,18\}$ (joined by the full lines). The points correspond to the values of the computational time needed for $N=10$ replications.}
\label{fig:time_PPMLE_pred}
\end{figure}

\textcolor{black}{In practice, we solve~(\ref{eqn:estimator}) by decomposing the multidimensional optimization problem into as many separate unidimensional optimization problems as there are pairs of variables. Each individual optimization problem is solved using a golden section search method, which is implemented in the \texttt{optimize} function of the \texttt{stats} R package. As a result, the execution time of the global optimization problem  is linear in $n$ and quadratic in $d$. This is illustrated in~Figure~\ref{fig:time_PPMLE_pred} in $\log_{10}-$scale.
  It can be seen that for a sample size of $n=100$ and $d=100$ variables, the estimation takes around $200$ s on our computer. Therefore, for instance, for a number of variables $d=1000$, the estimation should take close to $20 000s \sim 5.5h$. Note that this time was measured on a single x86-64 core using a non-parallelized version. A parallel option is available in $\texttt{heterocop}$. To approximate the final computational time, one can divide the estimated time by the used number of cores.}

\section{Application on real data}
\label{s:app}

We applied the proposed methodology to a data set from the International Cancer Genome Consortium (ICGC, see~\citet{ICGC}) regarding Breast Cancer in the United States with 990 donors. 

On each individual, several samples were collected on both healthy and tumoral tissue. Our variables of interest here are RNA-seq counts, protein abundance, and mutations. We kept for further analysis only the samples collected on tumoral tissue, and averaged the normalized protein expression and the RNA-seq counts per individual. The binary encoding was kept for the presence of the mutations for each individual. The initially selected variables prior to pre-processing contained:
\begin{itemize}
    \item RNA-seq counts for 20 501 genes observed on 939 individuals
    \item normalized protein abundance for 115 genes observed on 260 individuals
    \item presence of 107 249 mutations observed on 918 individuals 
\end{itemize}

\subsection{Data pre-processing}

First, the RNA-seq counts were normalized via the $\texttt{DESeq2}$ R package~\citep{deseq2}, which enables to study gene differential expression, and rounded to the next integer. The number of donors was reduced to 250 after intersecting the available data for all types of variables. In order to reduce the dimension while allowing a biological interpretation of the results, we restricted the analysis to the 108 genes found in common between the RNA-seq and protein data. Concerning the mutation data, we kept those present in at least two donors, reducing their number to 62. The genes associated to each mutation were then identified via the $\texttt{ensembldb}$ R package~\citep{ensembldb}. As there were only 4 common genes involving the mutations, RNA-seq and protein data, we decided to keep all 62 mutations. Our final dataset therefore contained 250 individuals  and 278 variables: 108 discrete RNA-seq counts, 108 continuous protein data and 62 binary mutations. Note that for the mutations, the proportion of ones has gone from 0.001 to 0.013 after data pre-processing.

\subsection{Results}

The copula correlation coefficients of model~(\ref{eqn:model}) were
estimated by~(\ref{eqn:estimator}) from the final dataset. For
comparison, we also estimated the Spearman's $\rho^S$ and Kendall's
$\tau$ coefficients.

\subsubsection{Comparison of the copula correlation coefficient with Spearman's $\rho^S$ and Kendall's $\tau$}

Figure~S7 from the Supplementary material shows an histogram of the estimates of the coefficients of Spearman's $\rho^S$, Kendall's $\tau$, and the proposed copula. We can see that the copula correlation coefficient seems to span the entire range of possible values from -1 to 1, while Spearman's $\rho^S$ and Kendall's $\tau$ seem to take smaller absolute values.

\begin{figure}[p]
  \centering
  \subfloat[copula versus Spearman]{\includegraphics[width=0.45\textwidth]{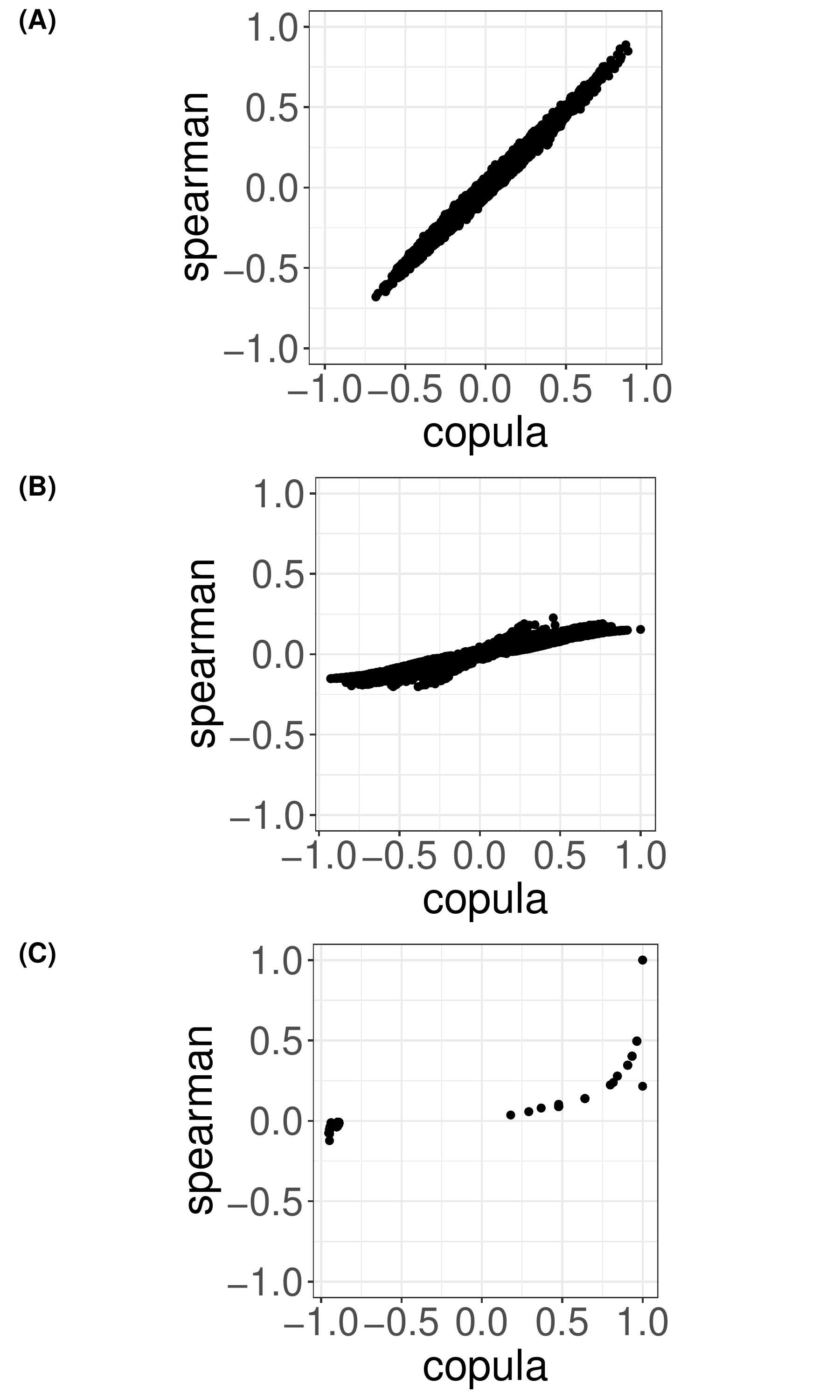}}
  \subfloat[copula versus Kendall]{\includegraphics[width=0.45\textwidth]{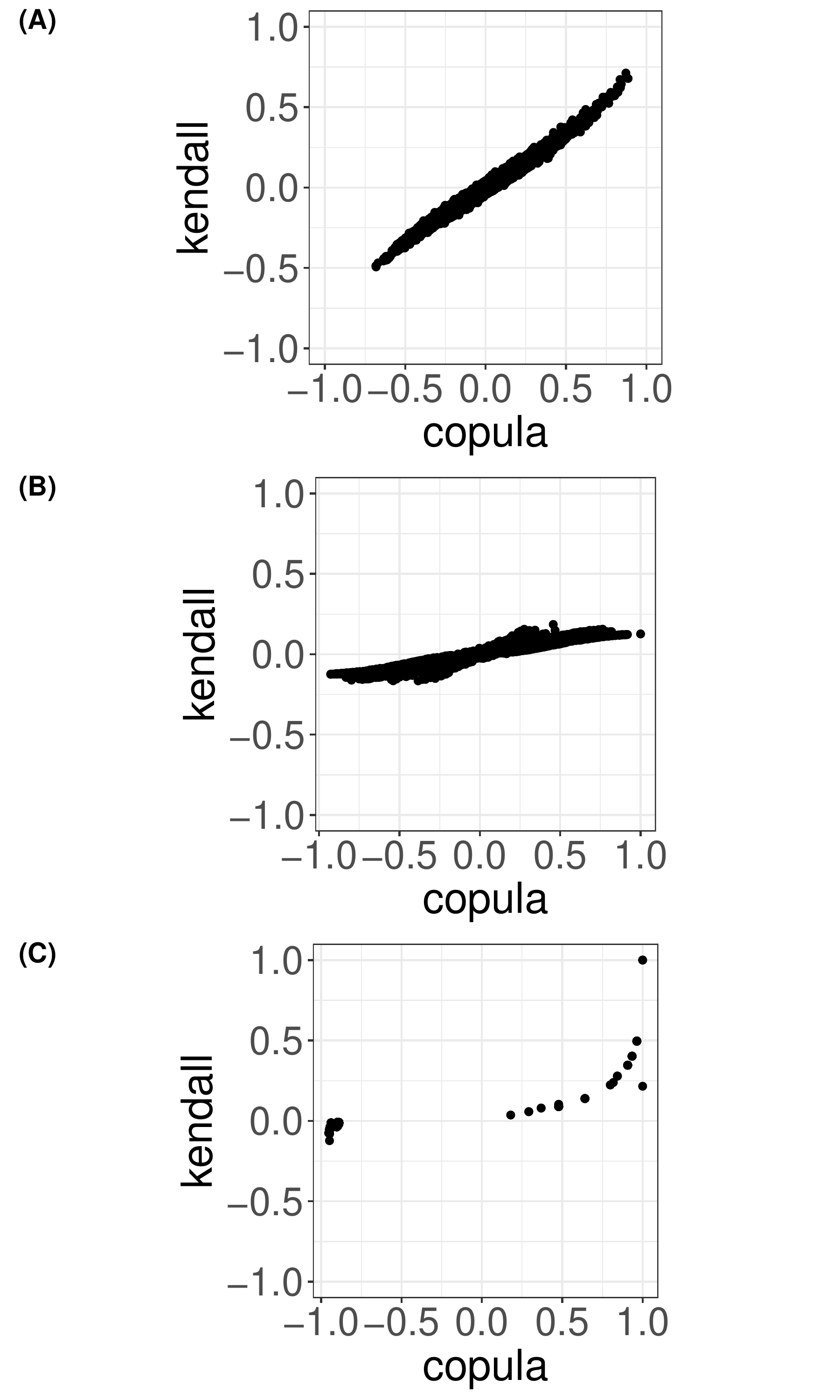}}
\caption{Copula correlation coefficient versus Spearman or Kendall coefficient for each combination of variable types:  continuous/continuous (A), binary/continuous (B) and binary/binary (C). The RNA-seq data have been grouped with the continuous data.}
\label{fig:comp}
\end{figure}

To understand the difference between Spearman's $\rho^S$, Kendall's $\tau$, and the copula correlation coefficient, we compare the estimates by type.  Remember that there are three variable types: discrete RNA-seq counts (D), continuous protein abundance (C) and binary mutations (B), and hence 6 possible combinations of types for each pair: DD, DC, DB, CC, CB, BB.    RNA-seq data, although discrete, have a large number of distinct values, which makes them nearly continuous. Hence we grouped the DD, DC and CC coefficient estimates, leaving three combinations CC (which also contains DD and DC), CB and BB.
A scatterplot is displayed for each of these combinations in   Figure~\ref{fig:comp}. Panel~A of Figure~\ref{fig:comp} confirms that the differences between Spearman's $\rho^S$ or Kendall's $\tau$ and the copula does not come from the combinations of types DD, DC and CC. 
We see that the differences are explained by the combinations involving the binary variables. The narrow range of Spearman's $\rho^S$ is explained by the fact that this coefficient applied to two Bernoulli variables with parameters $p_1$ and $p_2$ is bounded by $3p_1(1-p_2)$ in absolute value~\citep{mesfioui2022}.   

\subsubsection{Dependence relationships between the binary variables} Let $X_j$ denote the presence of the $j$th mutation in some individual ($X_j=1$ when the mutation is present and 0 otherwise) and let $p_j = \mathbb{P}(X_j=1)$ denote the Bernoulli parameter of $X_j$ ($j=1,\dots,62$). In the data all $p_j$ are less than 0.15. Thus the conditions of case~\eqref{item:Ber-Ber} of Proposition~\ref{prop:characterization} are satisfied by every pair of binary variables $(1-X_{j'},1-X_{j})$. Indeed $1-p_j+1-p_{j'}\ge 2-0.3 = 1.7 > 1$. Thus when the copula correlation coefficient is close to minus one, case~\eqref{item:Ber-Ber} of Proposition~\ref{prop:characterization} predicts that $1-X_j+1-X_{j'}>0$ and hence $X_{j}+X_{j'}\le 1$, that is, no two mutations can co-occur.  
Case~\eqref{item:Ber-Ber} of Proposition~\ref{prop:characterization} also predicts that when the copula correlation coefficient is close to one then $1-p_j<1-p_{j'}$ implies $1-X_j\le 1-X_{j'}$ and hence $X_j\ge X_{j'}$, that is, the rarest mutation cannot occur without the more common one. A look at the data confirms these predictions, see Table~S5 in the Supplementary material for an illustration.

\subsubsection{Network analysis}

From a set of estimates of copula correlation coefficients we can build a  network by linking highly dependent variables. More precisely, the network is a graph in which the nodes represent the variables and the edges the copula correlation coefficient estimates. One draws an edge between two variables if the absolute value of their copula parameter is greater than some chosen threshold.

\begin{figure}[p]
\includegraphics[width=\linewidth]{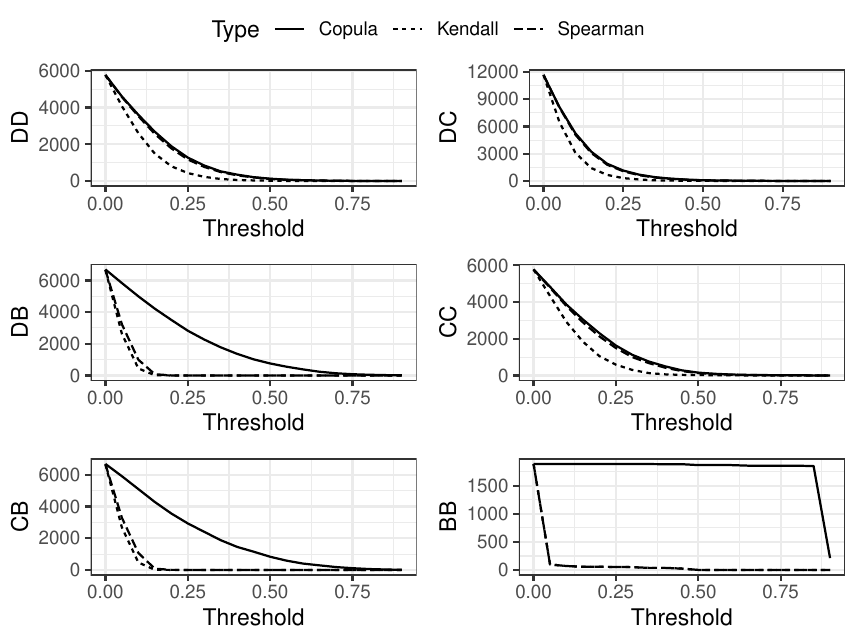}
\caption{Number of detected links by the copula,  Spearman, and Kendall coefficients, for threshold values ranging from 0 to 0.9, by combination of types where D stands for discrete (RNA-seq count data), C for continuous (protein data), B for binary (mutation data). In the panel BB the Spearman and Kendall curves coincide.}
\label{fig:number_links_tot}
\end{figure}

One can do the same with the estimates of the Spearman's $\rho^S$ or Kendall's $\tau$ coefficients, and comparison of the inferred networks by the three methods was investigated.
The number of detected edges as a function of the threshold, separately for each combination of data types (CC, CD, CB, DD, DB, BB) is depicted in Figure~\ref{fig:number_links_tot}. As illustrated in Figure~\ref{fig:number_links_tot}, for CC, CD and DD, the copula,  Spearman, and Kendall methods behave similarly and identify a similar number of links. When the binary mutation data are involved, however, the copula model detects more links than the Spearman or Kendall approach. This agrees with the previous remark that Spearman's $\rho^S$ between two binary variables in general cannot reach the endpoints of the interval $[-1,1]$. 

\begin{figure}[p]
  \centering
  \includegraphics[width=0.99\linewidth,
  trim=2cm 0cm 2cm 0cm, clip=true]{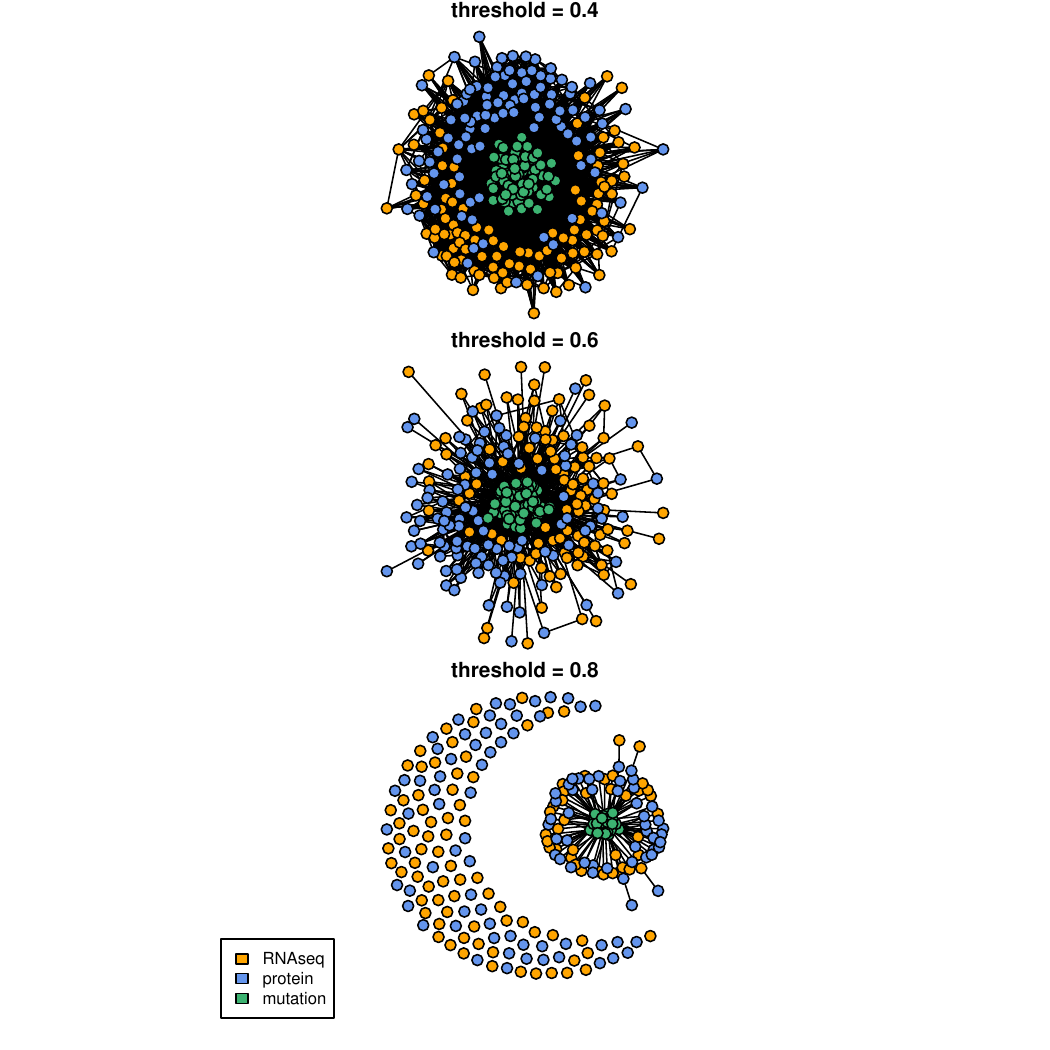}
    \caption{Copula correlation coefficient network for different threshold values. The nodes (variables)
      are colored by biological type (RNA-seq, protein, mutation). An
      edge is drawn between two nodes if the absolute value of the
      corresponding estimated correlation coefficient is above the
      threshold.}
  \label{fig:networks}
\end{figure}

Figure~\ref{fig:networks} presents the inferred networks obtained from model~(\ref{eqn:model}), for different threshold values. It can be noted that the strongest links are those among mutations, all at the center of the network. Around the mutations lie the genes (RNA-seq) and the proteins, tied to each other and to the mutations. We see that intra-type links (RNA-seq with RNA-seq, and proteins with proteins) tend to be strongest than inter-type links (RNA-seq with proteins), although this feature wanes as the threshold gets higher. By contrast, the networks based on Spearman $\rho^S$ or Kendall's $\tau$, depicted in supplementary Figure~S8, show mutations as peripheral nodes.

To check that the inferred links have biological meaning, we
arbitrarily selected the copula correlations' estimates above 0.7 (in
absolute value) and examined the nodes that had a degree (number of
connections) greater than 10.  Four variables were identified with
degrees 73, 73, 90, 83, corresponding to mutations MU4777833,
MU5153080, MU17289, MU5551967, respectively. The total number of
connections above the 0.7 copula correlation threshold pointing to
these mutations is then 319. Once we account for the uncertainty about
the copula correlations' estimates, this number remains as large as
216. To account for the estimates' uncertainties, we substracted to
each absolute value of the estimate 1.96 times the standard error of
the estimate.  The standard error was calculated as the square root of
the jackknife estimator of variance. See Section~H of the
Supplementary material.

The genes associated with the mutations MU4777833, MU5153080,
MU17289 and MU5551967, as identified by
$\texttt{ensembldb}$, are shown in Table~S6 in the
Supplementary Material.  We performed a literature search for each of
these genes, and they were all found to be involved in cancer
development.  Indeed, gene \textcolor{black}{\textit{ARHGEF11}} has been identified as playing a
key role in the migration and growth of invasive breast cancer
cells~\citep{Itoh2017}. Gene \textcolor{black}{\textit{SLC7A9}} belongs to the \textcolor{black}{\textit{SLC7}} family which
is known for its role in cancer cell
metabolism~\citep{Yan2022}. Similarly, gene \textcolor{black}{\textit{CDKN1B}} affects protein p27
which is linked to the production of breast cancer
cells~\citep{Cusan2018} and finally, \textcolor{black}{\textit{PQBP1}} is usually overexpressed in
breast cancer patients~\citep{Liu2024}.  The identified hubs of the
copula network therefore seem to highlight interesting mutations, that
were not identified with the Spearman or Kendall approach.

\section{Applicability}
\label{sec:applicability}

Our presented method, based on the semiparametric pairwise log-likelihood, is
truly general. We do not need any prior knowledge about the
(univariate) marginal distributions of the data, whether continuous or
discrete. In particular, we do not need to know the support of the
discrete data nor know or choose a function to transform the
continuous data to Gaussian data. We only need to decide which
variables shall be handled as discrete, and which as continuous. Of
course each categorical (discrete but not ordinal) variable needs to
be subjected to dummy encoding with respect to a reference
value.

From our experience the computation of the semiparametric pairwise
log-likelihood does not raise any particular computational issue. We
simply evaluate, for each pair, the bivariate log-density at each of
the observations of the given sample. As the objective function is
continuous it always has a solution in a closed interval containing
the true value of the copula correlation coefficient.

In practice the limitation of the method remains its computational
expense, which scales quadratically as the number of variables
increases.

\section{Discussion}
\label{s:discuss}

The joint analysis of heterogeneous data is a key methodological
topic, especially in the context of multi-omic analyses. We proposed
here an innovative approach based on the Gaussian copula that allows
to build correlation networks from various types of data (continuous
and discrete).  The proposed estimation method is based on a
semi-parametric pairwise likelihood for mixed-type data, with no
explicit assumption concerning the distribution of the marginals,
which makes it very flexible for biological data analysis. The
estimation procedure is implemented in a freely available R package
called \texttt{heterocop}.

We theoretically derived properties of the copula correlation
coefficients to make the link with the dependence relationships in the
observed data. In particular, we showed that a block-wise structure in
the copula correlation matrix is equivalent to block-wise mutual
independence in the observed data.  We characterized the lower and
upper extreme values of the copula parameter in terms of the observed
data when a Bernoulli distribution is involved, thus providing an
interpretation of the copula parameters.

In an extensive simulation study, we showed that under various
experimental designs the Gaussian copula correlation matrix was
estimated with a good accuracy with only dozens of observations even
for a large number of variables (several hundreds). We also showed
that it provided more accurate results than Kendall or Spearman
coefficients, especially for the analysis of binary data. This result
was also observed in the real data analysis regarding a breast cancer
study including binary mutation data.

Regarding the block-wise mutual independence property, it would be
interesting in a further work to propose a sound statistical procedure
to identify independent blocks in the data.

It would be of interest to devise theoretically grounded inferential
procedures to quantify the uncertainty about the copula parameters'
estimates.  Although the jackknife estimator of variance was applied
in Section~\ref{s:app}, its theoretical properties still need to be
investigated in the present context.  Another route would be to base
the inference on a normal approximation of the estimator's
distribution. We would need to establish the asymptotic normalilty of
the estimator and calculate the variance of the limit distribution. In
the present case this may not be straightforward because, unlike
standard semiparametric copula theory, the marginal cumulative
distribution functions can be discontinuous.  A possible way forward
would be to extend~\citet{ruymgaart_asymptotic_1974,Fermanian2024}.
This would open the gate to statistical testing and model selection.

Our focus was here on the correlation matrix estimation. In order to
obtain the direct links in the networks, the next step would be to
propose an estimation procedure for the precision matrix, using the
computational efficiency of the pairwise likelihood approach, with a
Lasso penalty to obtain a sparse network.

\section*{Supplementary materials}
\begin{itemize}
\item A PDF file containing supplementary numerical experiments,
  proofs, details of some mathematical calculations, and two
  resampling procedures for uncertainty quantification.
\end{itemize}
Supplementary materials are available online at
\footnotesize\url{https://hal.inrae.fr/hal-04847648}.\normalsize

\section*{Acknowledgments}
We thank the associate editor and two anonymous referees for their
careful reading and detailed comments that led to an improved version
of the manuscript. We are also grateful to the INRAE MIGALE
bioinformatics facility (MIGALE, INRAE, 2020. Migale bioinformatics
Facility,
\footnotesize\url{doi: 10.15454/1.5572390655343293E12}\normalsize)
\normalsize for providing
computing resources.

\bibliographystyle{abbrvnat}
\bibliography{biblio.bib}

\end{document}